\documentclass[%
 reprint,
nofootinbib,
 amsmath,amssymb,
 aps,
]{revtex4-2}

\usepackage{graphicx}
\usepackage{subcaption}
\usepackage{dcolumn}
\usepackage{bm}
\usepackage{xfrac}
\usepackage{hyperref}
\hypersetup{colorlinks=true,linkcolor=red,citecolor=blue,filecolor=blue,urlcolor=blue}
\usepackage{booktabs}
\usepackage{orcidlink}
\usepackage{physics}
\usepackage{multirow}
\usepackage{aasmacros}
\usepackage{url}
\usepackage{siunitx}
\usepackage{xspace}
\usepackage{comment}
\usepackage{textcomp}
\usepackage{tabularx}

\newcommand{\parfrac}[2]{\left(\frac{#1}{#2}\right)}
\newcommand{\faxe}{FAxE\xspace}

\begin{document}
\defcitealias{finke_modeling_2022}{Finke22}
\defcitealias{haardt_radiative_2012}{CUBA}

\graphicspath{{./}{figures/}}

\title{Sensitivity of a gigahertz Fabry-P\'erot resonator for axion dark matter detection}

\author{Jacob Egge$^1$\,\orcidlink{0000-0003-2852-3447}}
\email{jacob.mathias.egge@desy.de}
\author{Manuel Meyer$^2$\,\orcidlink{0000-0002-0738-7581}}
\email{mey@sdu.dk}
\affiliation{
 $^1$Deutsches Elektronen-Synchrotron DESY, Notkestr. 85, 22607 Hamburg, Germany\\
 $^2$CP3-Origins, University of Southern Denmark, Campusvej 55, DK-5230 Odense M, Denmark}

\date{\today}

\begin{abstract}

Axions are hypothetical pseudo-Nambu Goldstone bosons that could explain the observed cold dark matter density and solve the strong CP problem of QCD. 
Haloscope experiments commonly employ resonant cavities to search for a conversion of axion dark matter into photons in external magnetic fields.
As the expected signal power degrades with increasing frequency, this approach becomes  challenging at frequencies beyond tens of gigahertz. 
Here, we propose a novel haloscope design based on an open Fabry-P\'erot resonator. 
Operating a small-scale resonator at cryogenic temperatures and at modest magnetic fields should already lead to an unparalleled sensitivity for photon-axion couplings $g_{a\gamma} \gtrsim 3\times10^{-12}\,\mathrm{GeV}^{-1}$ at 35\,GHz. 
We demonstrate how this sensitivity could be further improved using graded-phase mirrors and sketch possibilities to probe benchmark models of the QCD axion.

\end{abstract}

\maketitle

\section{Introduction\label{section:Intro}}

The existence of dark matter has been firmly established through its gravitational interaction with baryonic matter~\cite{2005PhR...405..279B,2010ARA&A..48..495F}.
Prime candidates for dark matter are electrically neutral pseudoscalar axions and axionlike particles (henceforth collectively denoted as axions)~\cite{dine1983, preskill1983, abbott1983, arias2012},
originally introduced to solve the strong CP problem in QCD~\cite{pq1977,weinberg1978,wilczek1978}.
Such a QCD axion necessarily couples to gluons and its mass $m_a$ and coupling to photons $g_{a\gamma}$ are typically predicted to be proportional to each other~\cite{dfs1981,z1980,1979PhRvL..43..103K,svz1980}.
Recent theoretical models suggest a much wider parameter space of axions to be of interest. Examples are axion cogenesis models~\cite{2021JHEP...01..172C}, photophilic or photophobic axions~\cite{2021JHEP...06..123S,2018JHEP...09..028C}, or axions that could simultaneously explain dark matter and inflation, i.e., the exponential growth of the Universe shortly after the big bang~\cite{2017JCAP...05..044D}. 
Axions are also a common prediction of string theory~\cite{arvanitaki2010}.

Axions could be detected through their conversion to an electric field $\mathbf{E}$ in the presence of an external magnetic field $\mathbf{B}_e$, described by the Lagrangian density $\mathcal{L}_{a\gamma} = g_{a\gamma}\mathbf{E}\cdot\mathbf{B}_ea$, where $a$ is the axion field. The most sensitive searches have been carried out in the mass regime between $\SI{2}{\micro\eV} \lesssim m_a \lesssim \SI{50}{\micro\eV}$ utilizing the haloscope concept.
In a typical haloscope, a DC magnetic field is used to convert axions into an electric field with a frequency $f = m_a c^2 / h$ and wavelength $\lambda = c / f$ ($c$: speed of light in vacuum, $h$: Planck constant)~\cite{sikivie1983}.
Usually, a cavity with a resonance frequency equal to $f$ is used to increase the expected signal. 
Tunable microwave cavities with resonance frequencies at hundreds of MHz with large volumes $V$ immersed in magnetic fields $B_e$ of several Tesla have provided strong constraints on the axion-photon coupling~\cite[e.g.][]{2018PhRvL.120o1301D}. 
Dark matter axions would produce a narrow line at frequency $f$ with a line width $\Delta f / f \sim Q_a^{-1}$, where $Q_a \sim c^2 / \sigma_v^2 \sim 10^6$ is the axion quality factor and $\sigma_v$ the dark matter velocity dispersion in the dark matter halo of the Milky Way. 
The expected power from converted axions scales as~\cite{2020JCAP...03..066K}
\begin{equation}
 P_{\mathrm{sig}} = \frac{g_{a \gamma}^2 \rho_a}{m_a} \frac{\beta}{\beta + 1}\frac{Q_LQ_a}{Q_L+Q_a} V B_e^2 C,\label{eq:haloscope-power}
\end{equation}
where $Q_L$ is the loaded quality factor, i.e., the total energy stored in the cavity over the total power loss multiplied with the resonance frequency of the cavity, $C$ is a form factor, which measures the spatial overlap of the axion field with the cavity eigenmode, and $\beta$ is the receiver coupling strength. 
The axion dark matter density is given by $\rho_a$.
To maximize $C$, haloscope experiments typically aim to operate at low order cavity modes.
For example, the ADMX experiment employs a cylindrical cavity to search for axion dark matter at microwave frequencies and operates at the TM$_{010}$ mode yielding $C\sim 0.4$ with a loaded quality factor of $Q_L\approx5\times10^4$~\cite{2018PhRvL.120o1301D}. 
The diameter of the cylindrical cavity is, however, constrained to $\sim\lambda/2$ to still operate at the TM$_{010}$ mode.
The length of the cavity is \textit{a priori} independent of the wavelength but is in practice also limited to a few $\lambda$ due to unfavorable mode localization. Thus, the volume scales $V\propto \lambda^3$ and consequently the signal power diminishes at higher frequencies. 

Proposals to search for axion dark matter at tens and hundreds of GHz include the ORGAN experiment~\cite{2017PDU....18...67M} and CADEx~\cite{2022JCAP...11..044A}, which plan to operate a variety of resonant cavities of different dimensions.
In a different approach, haloscopes are equipped with dielectric disks. 
By carefully tuning the distance between the disks, a resonant enhancement of the axion-induced electric field can be achieved at higher-order modes while maintaining a high form factor.
Examples are the MADMAX and ADMX-Orpheus experiments, currently at a pathfinder stage~\cite{2019EPJC...79..186B,2022PhRvD.106j2002C}. 
Alternatively, the resonant approach can be abandoned altogether in favor of a broadband signal search using dish antennas~\cite{2013JCAP...04..016H}. 
This is the aim of, e.g., the planned BREAD experiment~\cite{2022PhRvL.128m1801L}. 

Here, we propose a different haloscope design to probe axion frequencies above 30\,GHz. 
Instead of a closed cavity, we suggest an open Fabry-P\'erot resonator (FPR) consisting of two curved aluminum mirrors at cryogenic temperatures within a strong magnetic field. For open resonators, the diameter of the mirrors is no longer constrained to the wavelength. Thus, the transverse area of our proposed Fabry-P\'erot Axion Experiment (\faxe) can reach $\sim10^3 \lambda^2$. This offers an intriguing potential to have larger volumes at higher frequencies while keeping a reasonably high form factor. 

The paper is organized as follows. In Sec.~\ref{sec:resonator-estimate}, we introduce the \faxe setup together with sensitivity estimates of the axion parameter space that can be probed. 
Details for the suggested mirror design and the resulting form and quality factors are presented in Sec.~\ref{sec:mirrors}. We discuss our results and provide an outlook on how to reach the sensitivity for the QCD axion in Sec.~\ref{sec:conclusions}.

\section{Experimental Design of \faxe}
\label{sec:resonator-estimate}

\begin{figure}[thb]
   \centering
    \includegraphics[width=.95\linewidth]{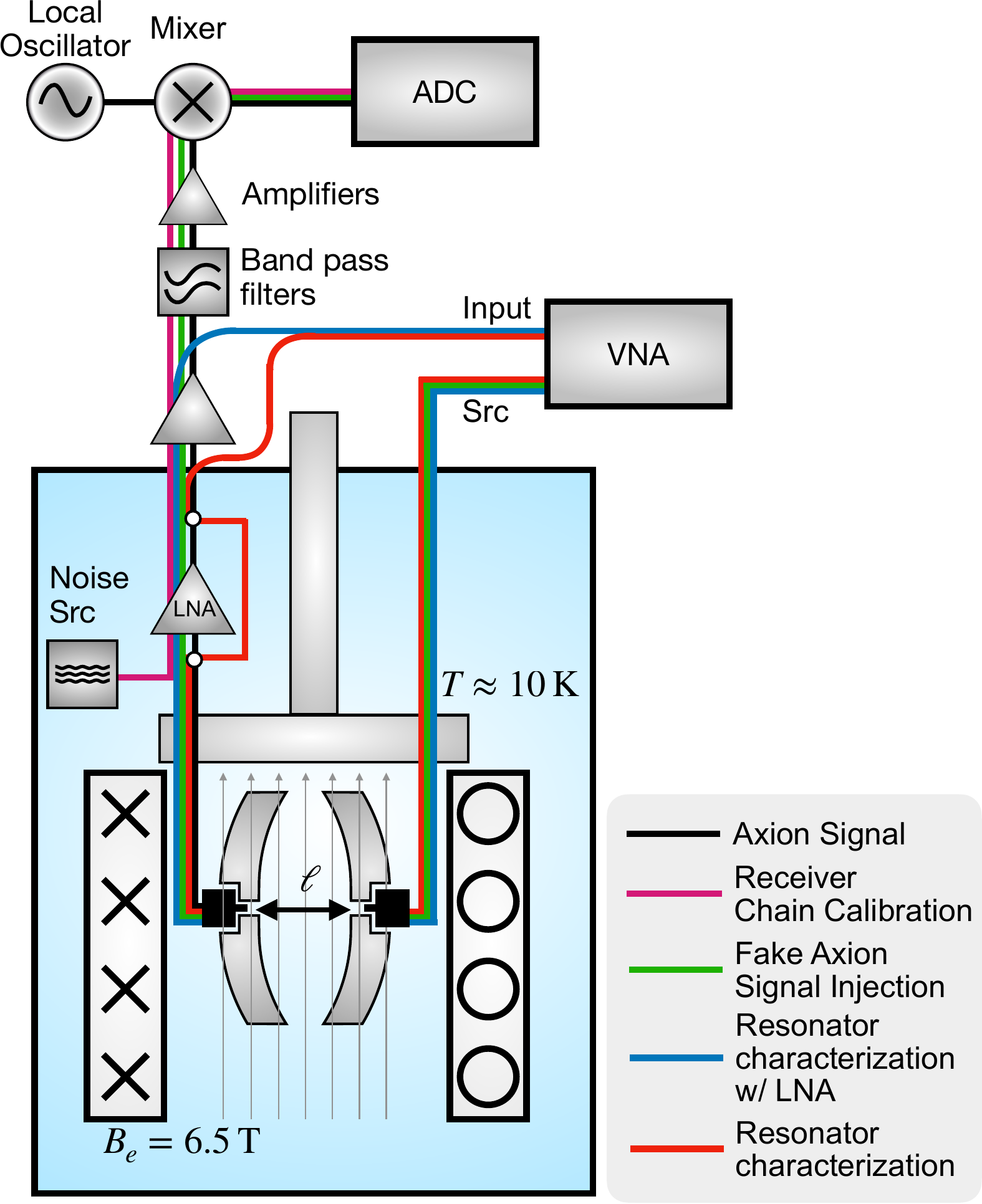}
    \caption{Diagram of the proposed Fabry-P\'erot haloscope. The different components are described in the text. Colored lines show the different calibration chains. 
    }
    \label{fig:layout}
\end{figure}

The basic idea of \faxe is to place an FPR with two 
curved aluminum mirrors inside a cryostat and a magnetic field, schematically shown in Fig.~\ref{fig:layout}.
The mirrors are placed at a distance $\ell$ from each other, which leads to resonance frequencies approximately at $q f_\mathrm{FSR}$, where $q$ is a nonzero integer and $f_\mathrm{FSR}$ is the free spectral range (FSR)~\cite{2016OExpr..2416366I}. 
In the following, we are only interested in the fundamental mode, as this leads to the largest overlap with the axion field, and set $q=1$.
The FSR is defined such that a monochromatic wave interferes constructively with itself after one cavity round trip, which takes the time $t_\mathrm{RC}$, so that
\begin{equation}
 f_\mathrm{FSR} = 1/t_\mathrm{RT} = c / (2\ell)\label{eq:fsr}
\end{equation}

Thus, for a certain axion frequency $f = 30\,\mathrm{GHz}~(m_a/1.24\times10^{-4}\,\mathrm{eV})$ resonating in the fundamental cavity mode, the length of the cavity needs to be tuned to 

\begin{equation}
    \ell = 5.0\,\mathrm{mm} \parfrac{30\,\mathrm{GHz}}{f} = 5.0\,\mathrm{mm} \parfrac{\lambda}{10\,\mathrm{mm}}.\label{eq:fsr_num}
\end{equation}

As a consequence, the resonator volume only scales linearly with the wavelength compared to $\lambda^3$ for cylindrical cavities. 
For photon-axion conversion, the magnetic field has to be oriented parallel to the mirror surfaces and the whole resonator needs to fit within the magnet bore inside the cryostat. 
Assuming a typical magnet bore with a diameter of 50\,mm, the mirror radius is limited to $r \approx 25\,$mm, resulting in a volume of $V\approx 10\,\mathrm{cm}^3\,(r / 25\,\mathrm{mm})^2 (\ell / 5\,\mathrm{mm})$.
At $\SI{30}{\giga\hertz}$, this corresponds to a volume $V\sim 10\lambda^3$, which is comparable to  
volumes achieved by 
other haloscopes sensitive in a similar frequency range
(e.g., the ORGAN Phase 1b experiment operates a cavity in the TM$_{110}$ mode at a resonance frequency around $\sim 26\,$GHz and has a volume of $V\sim 8\lambda^3$ with $C = 0.65$~\cite{2024PhRvL.132c1601Q}).

The mirror design and the optimal radius of curvature will be discussed  in Sec.~\ref{sec:mirrors}. There is a tradeoff between the loaded quality factor $Q_L$, which depends on the radius of curvature of the mirrors, and maintaining a high form factor $C$.
We find an optimal radius of curvature that leads to a small field leakage outside of the open resonator and a resonant enhancement at its center with $Q_L = 2.7\times 10^4$ (assuming the conductivity of aluminum at $\SI{4}{\kelvin}$) and $C=0.28$ at $f=35\,$GHz ($m_a = 1.45\times10^{-4}\,\mathrm{eV}$).
Using  Eq.~\eqref{eq:haloscope-power} and assuming a 6.5\,T external magnetic field, we therefore expect an axion signal power in our resonator of 

\begin{widetext}
\begin{align}
    P_\mathrm{sig} &= 3.7\times10^{-23}\,\mathrm{W} \parfrac{\beta}{\beta+1}\parfrac{Q_LQ_a / (Q_L + Q_a)}{2.7\times10^4}\parfrac{1.45\times10^{-4}\,\mathrm{eV}}{m_a}\parfrac{g_{a\gamma}}{10^{-12}\,\mathrm{GeV}^{-1}}^{2} \nonumber\\
    & \times \parfrac{B_e}{6.5\mathrm{T}}^2 \parfrac{C}{0.28}\parfrac{V}{8.57\,\mathrm{cm^3}}\parfrac{\rho_a}{0.45\,\mathrm{GeV}\,\mathrm{cm}^{-3}}.\label{eq:signal-power}
\end{align}
\end{widetext}

The signal is extracted through an iris at the mirror center whose size is tuned so that a coupling factor of $\beta\sim 1$ is achieved (see  Sec.~\ref{sec:mirrors}).
It is then transmitted through a low-loss superconducting coaxial cable  to a low noise amplifier (LNA), as shown in Fig.~\ref{fig:layout}. 
Similar to Ref.~\cite{2022SciA....8.3765Q}, the LNA should be placed just above the magnet bore to guarantee a low system temperature, $T_\mathrm{sys}\lesssim10\,$K. 
Outside the cryostat, the signal is further cleaned and amplified through a series of amplifiers and bandpass filters.
Just as in other haloscope experiments, the heterodyne detection scheme can be used in \faxe to search for a potential axion signal. 
The signal from the resonator is mixed with a periodic signal from a local oscillator to generate a beat note within the bandwidth of the analog-to-digital converter (ADC). The recorded spectra can then be analyzed for a potential axion line. 

In Fig.~\ref{fig:layout}, we also show a number of calibration chains to determine the sensitivity and response of the resonator inspired by previous haloscopes~\cite[e.g.,][]{2018PhRvL.120o1301D,2022SciA....8.3765Q}. 
The FPR itself will be characterized by sending in signals at different frequencies by means of a vector network analyzer (VNA). 
The signal is generated at the source (Src) port and measured at the input port. 
In this way, one can measure the exact resonance frequency and quality factor \textit{in situ}. 
With RF switches we can bypass the LNA to determine its effect on the signal. 
A signal generator (for example from a VNA) can be used to generate fake axion signals to assess the response of the full setup to a potential signal.
A calibrated cryogenic noise source with tunable temperature inside the cryostat serves as a calibrator of the receiver chain using the Y-factor method. The form factor can be assessed via bead-pull measurements which also open up the possibility to directly measure $P_\mathrm{sig}$ via the reciprocity approach developed in Refs.~\cite{Egge:2022gfp,Egge:2023cos,MADMAX:2024jnp}.

The achievable signal-to-noise ratio $S/N$ of the experimental setup for an observation time $\Delta t$ is determined through the Dicke equation,
\begin{equation}
   \frac{S}{N}   = \frac{P_{\mathrm{sig}}}{k_BT_{\mathrm{sys}}} \sqrt{\frac{\Delta t}{\Delta f}},
   \label{Eq:dicke_radiometer}
\end{equation}
with $\Delta f = f/Q_a$. 
Plugging in the signal power $P_\mathrm{sig}$ from Eq.~\eqref{eq:signal-power}, we obtain the photon-axion coupling $g_{a\gamma}$ that we can probe with \faxe for a given $S/N$ ratio and observation time, 
\begin{widetext}
\begin{align}
    g_{a\gamma} = 3.74\times10^{-12}\,\mathrm{GeV}^{-1} &
    \parfrac{S/N}{3}^{\frac{1}{2}}
    \parfrac{T_\mathrm{sys}}{10\,\mathrm{K}}^{\frac{1}{2}}
    \parfrac{1\,\mathrm{d}}{\Delta t}^{\frac{1}{4}}
    \parfrac{f}{35\,\mathrm{GHz}}^{\frac{3}{4}} \nonumber\\
&   \times
    \parfrac{1+1/\beta}{2}^{\frac{1}{2}}
    \parfrac{2.7\times10^4}{Q_LQ_A/(Q_L + Q_a)}^{\frac{1}{2}}
    \parfrac{6.5\mathrm{T}}{B_e} \nonumber\\
&   \times
    \parfrac{0.28}{C}^{\frac{1}{2}}
    \parfrac{10\,\mathrm{cm^3}}{V}^{\frac{1}{2}}
    \parfrac{0.45\,\mathrm{GeV}\,\mathrm{cm}^{-3}}{\rho_a}^{\frac{1}{2}}.\label{eq:g-sens}
\end{align}
\end{widetext}

\begin{figure}[thb]
    \centering
    \includegraphics[width=0.9\linewidth]{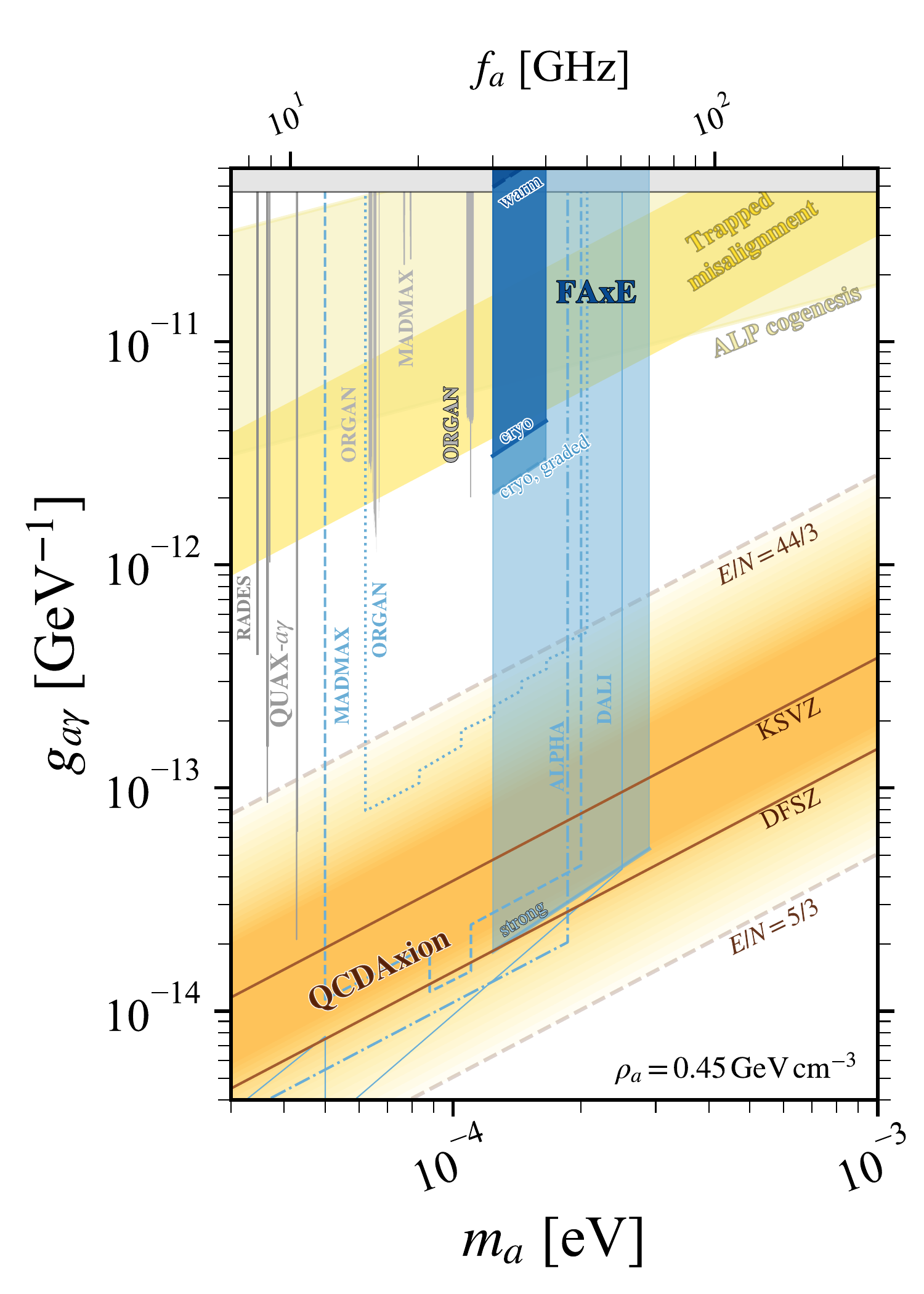}
    \caption{The projected \faxe sensitivity for different considered configurations (see Table~\ref{tab:configs}) in terms of axion mass and photon-axion coupling for an assumed local axion dark matter density of $\rho_a = \SI{0.45}{\giga\eV\per\cubic\centi\meter}$ (in accordance with, e.g., Ref.~\cite{2021RPPh...84j4901D}).
    We also show the KSVZ and DFSZ QCD axion benchmark models~\cite{1979PhRvL..43..103K,svz1980,dfs1981,z1980} for different ratios $E/N$ of the electromagnetic and QCD anomaly coefficients, the ALP cogenesis model prediction~\cite{2021JHEP...01..172C}, as well as the predictions for  trapped misalignment~\cite{2021arXiv210201082D}.
    Also shown are existing exclusions from 
    MADMAX~\cite{2024arXiv240911777A}, ORGAN~\cite{2017PDU....18...67M, 2022SciA....8.3765Q,2024PhRvL.132c1601Q,2024arXiv240718586Q}, RADES~\cite{2021JHEP...10..075A,2024arXiv240307790A}, QUAX \cite{2024PhRvD.110b2008R}, as well as from the CAST experiment~\cite{CAST:2024eil}.
    Future sensitivities of the haloscope experiments ALPHA~\cite{2023PhRvD.107e5013M}, DALI~\cite{2024PhRvD.109f2002D}, MADMAX~\cite{2020arXiv200310894B}, and ORGAN~\cite{2017PDU....18...67M} are shown as blue dashed-dotted, solid, dashed, and dotted lines, respectively.
    }
    \label{fig:m-vs-g}
\end{figure}

The resonance frequency and therefore the axion mass to which \faxe is sensitive can be tuned by changing the distance between the mirrors.
For example, a tuning range of $2.1\,\mathrm{mm} \leqslant \ell \leqslant 5\,\mathrm{mm}$ would correspond to a scan range between between 30\,GHz and 70\,GHz.
However, a more narrow tuning range between 30\,GHz and 40\,GHz appears more feasible at least in a first stage of \faxe given the frequency ranges of typical commercially available VNAs.
The possible signal scan rate $df/dt$ is given by~\cite{Irastorza2018, 2024PhRvD.110d3022C}, 
\begin{equation}
\label{eq:scan_rate}
    \frac{df}{dt} =  Q_aQ_L\left(\frac{1}{(S/N)}\frac{1}{k_B T_\mathrm{sys}}\frac{\beta}{\beta + 1}B_e^2 C V \frac{\rho_a g_{a\gamma}^2}{m_a}\right)^2.
\end{equation}
To reach our baseline sensitivity at cryogenic temperatures 
given in Eq.~\eqref{eq:g-sens} over a frequency range from $\SI{30}{\giga\hertz}$ to $f_\mathrm{max}=\SI{40}{\giga\hertz}$ using spherical mirrors and setting $\beta = 1$, $S/N = 3$, 
and $\Delta t = \SI{1}{\day}$, a total measurement time of $t_\mathrm{tot}\sim\SI{300}{\day}$ would be required (we refer to this baseline as the \emph{cryo} configuration).
The parameter space that can be probed within this frequency or mass interval is shown in Fig.~\ref{fig:m-vs-g}, where we also show sensitivities for different configurations with parameters given in Table~\ref{tab:configs} and motivated in the following sections.\footnote{
The data for the \faxe sensitivity shown in Fig.~\ref{fig:m-vs-g} as well as the data for Figs.~\ref{fig:FoM_gauss}, \ref{fig:scan_range}, \ref{fig:GP_mirror_profile}, and \ref{fig:GP_e_field} are available online~\cite{dataset_zenodo}.
}

\begin{table*}[thb]
    \centering
    \begin{tabular}{c|ccccccccc}
    \hline
    \hline
    Configuration         & Mirror design & $B_e~(\SI{}{\tesla})$ & $r~(\SI{}{\milli\meter})$ &  $Q_L$ &  $C$ &   $T_\mathrm{sys}~(\SI{}{\kelvin})$ & $f_\mathrm{max}~(\SI{}{\giga\hertz})$ &  $\Delta t~(\SI{}{\day})$ &  $t_\mathrm{tot}~(\SI{}{\day})$ \\
    \hline
	Warm              & spherical         & 6.5 &    25 &    2680 &  0.32 &   300 &    40 &  1.00 &        0.7 \\
	Cryo              & spherical         & 6.5 &    25 &   26800 &  0.28 &    10 &    40 &  1.00 &        675 \\
	Cryo, graded      & graded-phase      & 6.5 &    25 &   28750 &  0.57 &    10 &    40 &  1.00 &        675 \\
	Strong            & graded-phase, SC  & 9.4 &   400 & $10^6$ &  0.57 &  6.72 &    70 &  0.80 &       3522 \\
    \hline
    \end{tabular}
    \caption{Experimental parameters for the different proposed configurations of \faxe.}
    \label{tab:configs}
\end{table*}

\section{Optimizing the mirror design}
\label{sec:mirrors}

As seen from Eq.~\eqref{eq:haloscope-power}, 
the signal power scales as $P_{\mathrm{sig}}\propto Q_L C V$. 
As the available volume is often constrained by magnet and cryostat size, 
we are left to optimize both the quality factor $Q_L$ and form factor $C$. 
Maximum power can be extracted if the resonator is critically coupled to the 
receiver chain, i.e. $\beta=1$, and it is sufficient to optimize the unloaded quality factor $Q_0 = Q_L (1 + \beta)$. 
The unloaded quality factor can in turn be decomposed into individual $Q$-factors from different loss channels. 
For now, we only concern ourselves with losses due to the finite size of the mirror that set $Q_{\mathrm{diff}}$ and group all other losses into a single $Q_{\mathrm{other}}$ which include, e.g., conductive and scattering losses.
These depend largely on the mirror material and surface polish and are independent of the overall mirror geometry. 
The unloaded $Q$-factor is then
\begin{equation}
    \frac{1}{Q_0}  =\frac{1}{Q_{\mathrm{diff}}} + \frac{1}{Q_{\mathrm{other}}}.
\end{equation}
For our optimization of the mirror design, 
we define the figure of merit ($\mathrm{FoM}$) for peak sensitivity as 
\begin{equation}
    \mathrm{FoM} = Q_{0} C / Q_{\mathrm{other}},
\end{equation}
which can range between zero and one and allows us to compare mirror designs with  different quality factors. 

\subsection{Spherical mirror}
\label{sec:spherical_mirror}
In a common FPR, the mirrors follow a spherical profile. For simplicity, we assume a symmetric design in which the two mirrors have an identical shape. 
The mirror positions along the $z$-axis are at 
$z_m = \pm \ell / 2$ with the origin at the center between the mirrors.
The fundamental mode of an FPR can be described by a Gaussian beam, which is defined by its waist radius $w_0$ at which its amplitude has decreased by $1/e$. 
The waist radius also sets the radius of curvature of the phase fronts of the Gaussian beam, assumed to propagate along the $z$ direction,
\begin{equation}
    R(z) = z \left(1 + \left(\frac{\pi w_0^2}{\lambda z}\right)^2\right).
    \label{Eq:R_gauss}
\end{equation}

When the radius of curvature of the mirrors at position $z_m$ equals $R(z_m)$, the resonating eigenmode is approximately a Gaussian beam with corresponding waist radius. The larger $w_0$, the better the beam fills the resonator volume and the form factor $C$ increases. 
However, at the same time diffractive losses increase as well, reducing $Q_{\mathrm{diff}}$. 
Clipping due to the finite size of the mirrors can in turn feedback on the shape of the eigenmode, affecting $C$.
As a consequence, the resulting eigenmodes are best described numerically and we turn to simulations using the finite element method (FEM) to find the optimal mirror profile.

\begin{figure}
    \centering
    \includegraphics[width=\linewidth]{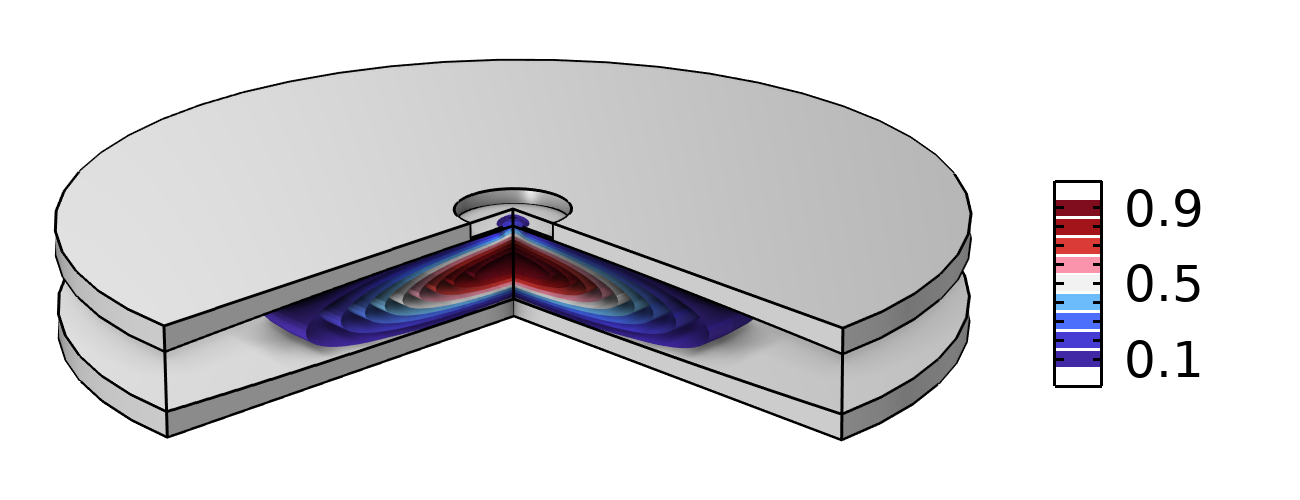}
    \caption{FEM simulation of an FPR with aperture. The contours show the normalized electric field of the fundamental resonating eigenmode.
    }
    \label{fig:comsol_spherical}
\end{figure}

We use COMSOL Multiphysics\textsuperscript{\textregistered} and its electromagnetic eigenmode solver from which we obtain $Q_0$ and $C$. We exploit the azimuthal symmetry of the setup to minimize computational demand. The linearly polarized electromagnetic fields can be decomposed into left and right circularly polarized fields which conform to azimuthal symmetry~\cite{2021JCAP...10..034K}.\footnote{See also \url{https://www.comsol.com/blogs/2d-axisymmetric-model-conical-horn-antenna}.}
Figure \ref{fig:comsol_spherical} shows the FPR with its fundamental eigenmode and an optional aperture to extract a signal. The electric field of the eigenmode is predominantly polarized parallel to the surface of the mirrors along which the external magnetic field would be applied. Small longitudinal components (perpendicular to the mirrors' surfaces) do not couple to such a external magnetic field and consequently reduce the form factor by  $\sim2\%$.
This effect is taken into account in the calculations of form factors given below.

The mirror size is set to $r = \SI{25}{\milli\meter}$ while the radius of curvature is set to be equal to $R(z_m)$ according to Eq.~\eqref{Eq:R_gauss} for different waist radii $w_0$.
Following Eq.~\eqref{eq:fsr}, the distance between mirrors is set to $\lambda / 2$ at $f=\SI{35}{\giga\hertz}$. 
The actual resonance frequency is around $\SI{36}{\giga\hertz}$ for the waist radii considered.

The mirror conductivity is set to either approximate aluminum at room temperature, $\sigma_{\mathrm{RT}} =\SI{5e7}{\siemens\per\meter}$, or at $\SI{4}{\kelvin}$, $\sigma_{\mathrm{4K}} =\SI{5e9}{\siemens\per\meter}$ which roughly corresponds to aluminum's improved conductivity at cryogenic temperatures~\cite{PAWLEK196614}. 
As conductive losses are the only other loss mechanism implemented in the simulation, they set the level of $Q_{\mathrm{other}}$, namely $Q_{\mathrm{other}}=5800$ and $Q_{\mathrm{other}}=58000$, respectively.
These values are extracted from the COMSOL simulation in the limit $w_0 \to 0$, where the conductive losses dominate. 
We refer to these two cases as the \emph{warm} (room-temperature) and \emph{cryo} benchmarks (see also Table~\ref{tab:configs}).

\begin{figure}
    \centering
    \includegraphics[width=\linewidth]{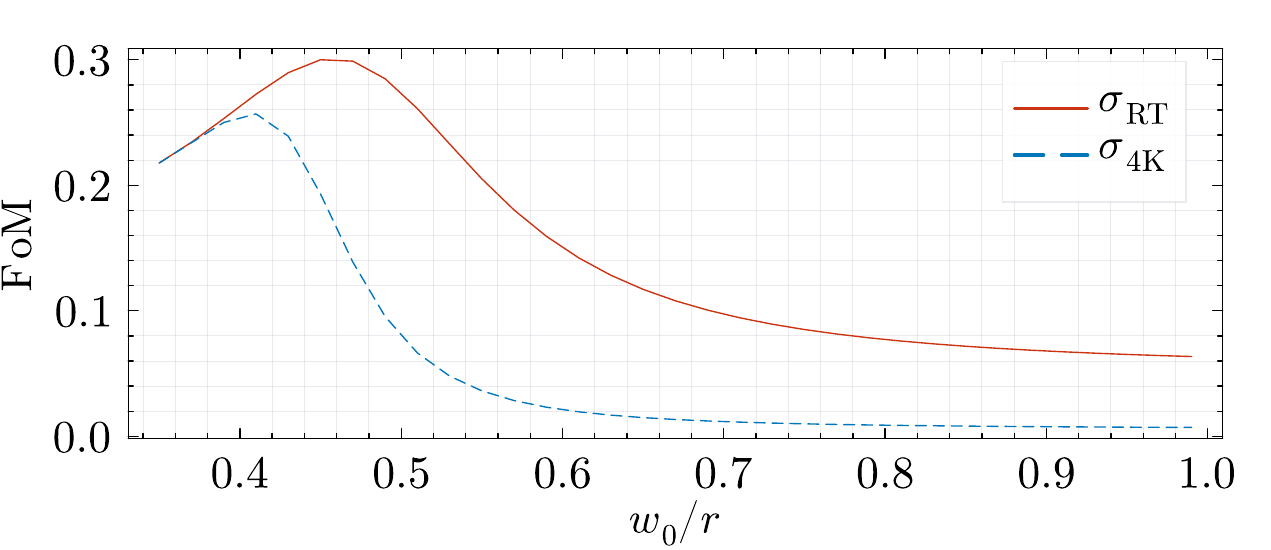}
    \caption{The figure of merit as a function of waist radius for a Gaussian beam FPR for different values of $Q_{\mathrm{other}}$, stemming from the different conductivities of aluminum at different temperatures.
    }
    \label{fig:FoM_gauss}
\end{figure}

The resulting FoM is shown in Fig.~\ref{fig:FoM_gauss} as a function of $w_0/r$.
For the room-temperature benchmark, the maximum occurs at $w_0/r \sim 0.45$ at which $Q_0 \sim 5360$ and $C \sim 0.32$. For the cryogenic benchmark, the optimum occurs at slightly smaller waist, $w_0/r_d \sim 0.41$ with  $Q_0 \sim 53600$ and $C \sim 0.28$.
The shift towards smaller $w_0/r_d$ is because, for a Gaussian beam, $Q_{\mathrm{diff}}$ increases approximately exponentially whereas $C$ only decreases quadratically for $w_0 \rightarrow 0$. This makes a tradeoff towards smaller waist radii worthwhile as long as diffractive losses are dominating, i.e. $Q_{\mathrm{diff}} < Q_{\mathrm{other}}$. 
Note that the signal power scales as $\mathrm{FoM}\cdot Q_\mathrm{other}$, so that we expect a larger axion signal at cryogenic temperatures.

\subsection{Frequency scan}

The axion-induced signal can be coupled to a receiver chain via a small circular aperture in one of the mirrors. The coupling factor $\beta$, which is the ratio between extracted power and dissipated power, depends on the size of aperture, its depth, and unloaded $Q$-factor $Q_0$. From Eq.~\eqref{eq:haloscope-power}, the maximum axion-induced power is extracted for critical coupling, $\beta=1$. 
By including a small aperture in our FEM simulation, we can calculate the required aperture parameters for critical coupling. In the cryogenic benchmark, the resulting aperture radius is $\sim 0.09 \lambda$ with a depth of $\sim 0.016 \lambda$.\footnote{
An overcoupled setup with $\beta = 2$, which is suggested to be the preferable choice for continuous data runs (see footnote 43 in Ref.~\cite[][]{2017PhRvD..96l3008B}), would also be possible and would require a similar optimization as performed here. The aperture would have to be larger, which would in turn affect the form factor. Our current choice for $\beta = 1$ maximizes the sensitivity of the \faxe. 
}

\begin{figure}
    \centering
    \includegraphics[width=\linewidth]{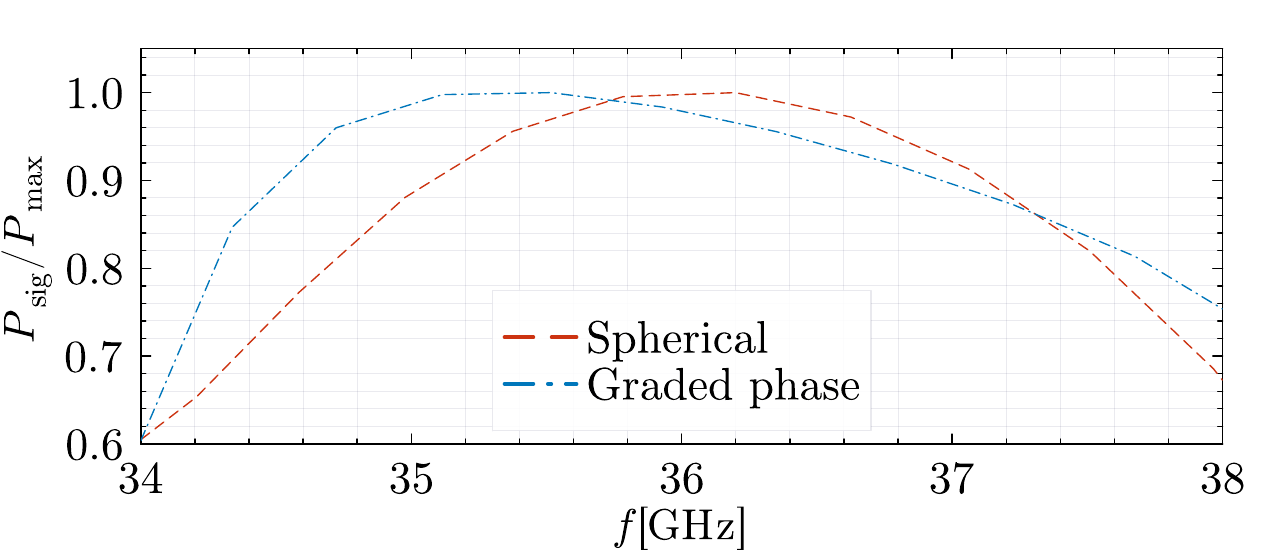}
    \caption{Signal power of a simulated FPR with fixed mirror geometry and aperture parameters as a function of resonance frequency which is tuned by changing the distance between the spherical and 
    graded-phase mirrors (see also Sec.~\ref{sec:graded-phase-mirrors}).
    Mirror and aperture geometry were optimized to $f = \SI{36}{\giga\hertz}$.
    }
    \label{fig:scan_range}
\end{figure}

Scanning over possible axion masses necessitates tuning of the resonance frequency. This can be achieved by changing the distance between the mirrors. However, as the mirrors' curvature and aperture parameters can practically only be optimized at a fixed resonance frequency, a reduction in axion-induced power is expected. Figure~\ref{fig:scan_range} shows the relative reduction in $P_{\mathrm{sig}}$ as a function of resonance frequency for the cryogenic benchmark.
The reduction is less than 10\% in a frequency range $\SI{2.2}{\giga\hertz}$ and is mainly from changes in $\beta$. 
However, to scan the entire frequency range from $\SIrange{30}{40}{\giga\hertz}$, different mirrors with different aperture parameters should be used.

\subsection{Graded-phase mirrors}
\label{sec:graded-phase-mirrors}

The tradeoff between diffractive losses and form factor leads to $\mathrm{FoM}\sim 0.25$ for an ordinary FPR with spherical mirrors.  This indicates that there is an untapped potential to further increase the FoM if one could maximize $C$ and $Q_{\mathrm{diff}}$ independently. Considering that the scan rate is $\propto C^2$, an improvement in $C$ without compromising $Q_0$ is even more desirable. One solution is to consider different transverse mode profiles that fill the area more efficiently and drop to zero more rapidly, minimizing diffractive losses. In theory, a boxlike profile would be best. A class of functions that describe both a Gaussian and boxlike profile are so-called super-Gaussians,
\begin{equation}
    E(x,y) = \exp\left(-\left(\frac{x^2 + y^2}{w_0^2}\right)^{N}\right).
\end{equation}

where $N$ is an integer. For $N\rightarrow\infty$, the profile approaches a box function of radius $w_0$. To have a resonating mode follow a super-Gaussian profile, the curvature of the mirrors needs to be adapted. Graded-phase mirrors have been demonstrated in the optical regime with super-Gaussian profiles of orders up to $N=6$~\cite{Blanger1992,Gerber2003}.
To our knowledge, they have not been attempted at microwave frequencies.

The method to design the profile of the mirrors is outlined in Ref.~\cite{Belanger:91} in which analytical calculations were employed. We instead use a separate FEM simulation. First, the desired transverse profile of the electric field at $z=0$ is chosen and propagated in axial direction in the absence of any resonator structure. Diffraction causes the electric field profile to spread and acquire a nontrivial phase front. The coordinates at which the phase of the electric field has changed by $\pi/2$ is then extracted from the simulation. If the mirrors of an FPR follow these extracted coordinates, the resonating eigenmode will in turn follow the desired transverse profile. 

\begin{figure}
    \centering
    \includegraphics[width=\linewidth]{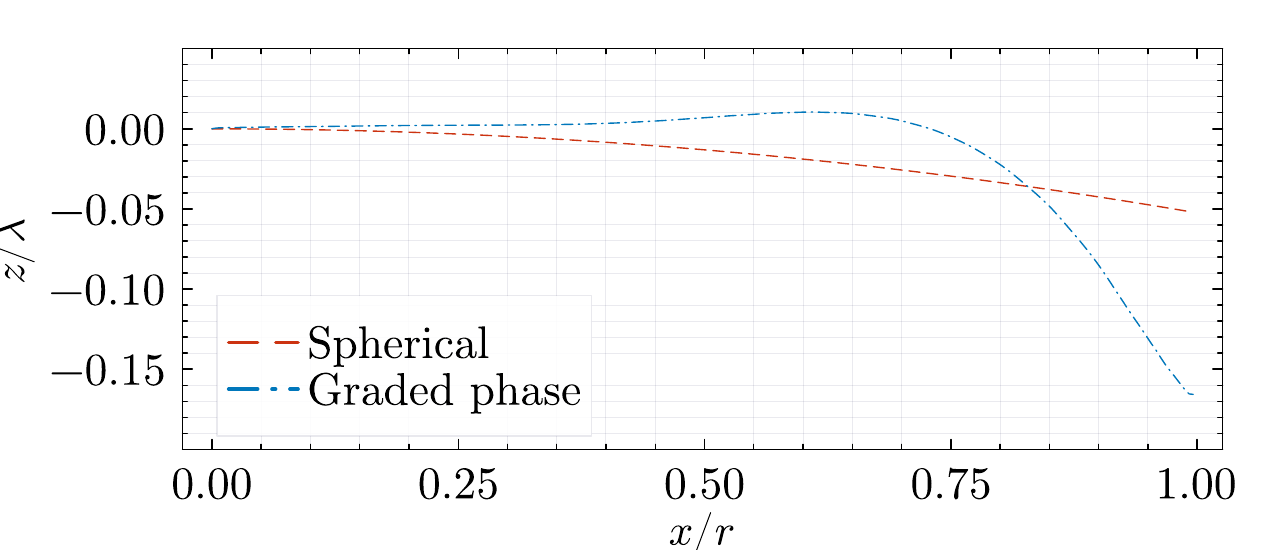}
    \caption{Profile of a graded-phase mirror ($N=3$, dash-dotted line) compared to a traditional spherical mirror (dashed line). Both are calculated using $w_0=0.73 r$.}
    \label{fig:GP_mirror_profile}
\end{figure}

\begin{figure}
    \centering
    \includegraphics[width=\linewidth]{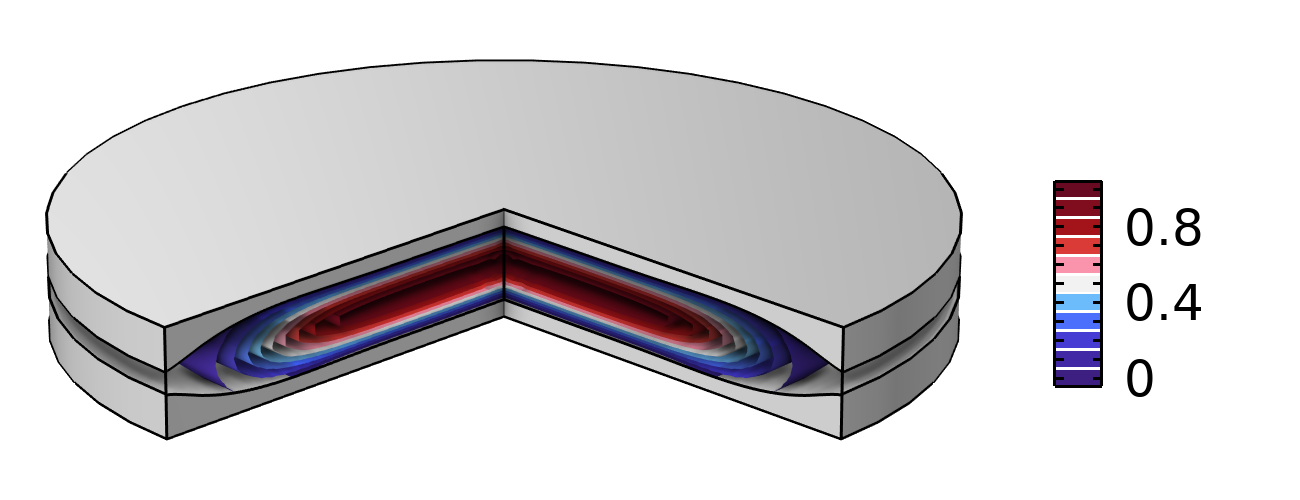}
    \caption{Same as Fig.~\ref{fig:comsol_spherical} but with the FEM simulation including graded-phase mirrors.}
    \label{fig:comsol_graded_phase}
\end{figure}

\begin{figure}
    \centering
    \includegraphics[width=\linewidth]{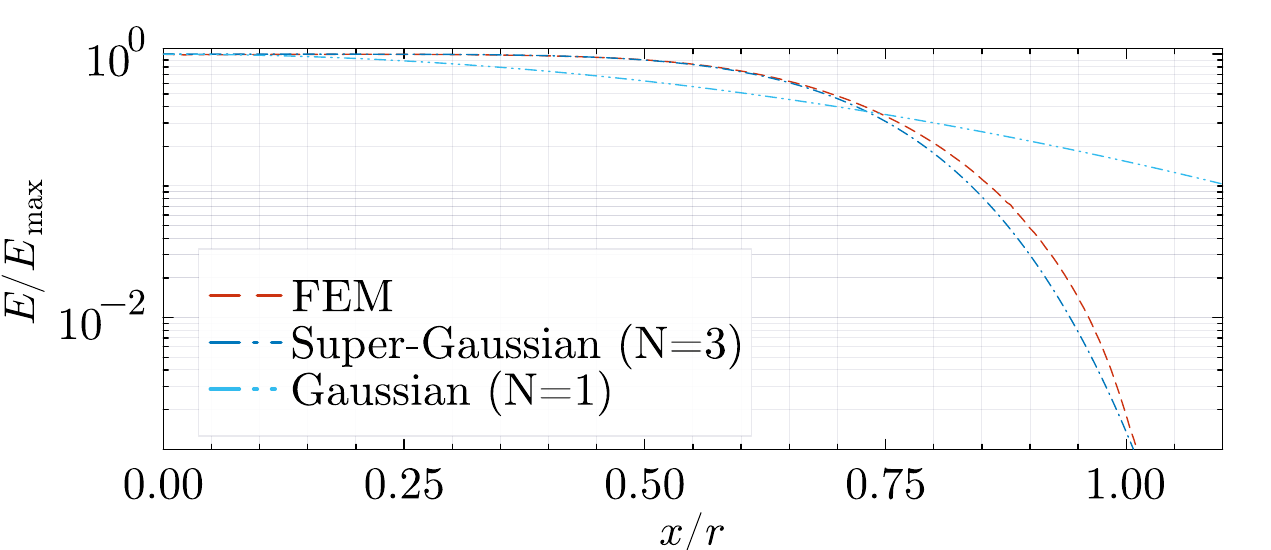}
    \caption{Transverse slice of the normalized electric field at $z=0$ from a FEM simulation of a graded-phase mirror FPR (dashed line). It closely follows the desired super-Gaussian profile (dash-dotted line). A Gaussian mode (dash-dot-dotted line) of  identical $w_0$ is shown for comparison.}
    \label{fig:GP_e_field}
\end{figure}

We apply the procedure described above to super-Gaussian profiles of different order $N$ and waist radius $w_0$. The larger $N$, the more boxlike the transverse mode profile becomes. However, the mirror profile also becomes increasingly complex and thus we select $N=3$ as a realistic compromise.
Figure \ref{fig:GP_mirror_profile} shows the calculated mirror profile for $N=3$ and $w_0=0.73 r$ which yields the optimal FoM for the mirror conductivity and size of the cryogenic benchmark.
Compared to the spherical mirror profile, also shown, the graded-phase mirror remains mostly flat in the center and confines the resonating mode towards the edges. The FPR with graded-phase mirrors (without aperture) is then simulated once more. Figure~\ref{fig:comsol_graded_phase} shows the geometry and resulting eigenmode. A transverse slice at $z=0$ of the electric field can be seen in Fig.~\ref{fig:GP_e_field}. The simulated resonating mode closely follows the desired profile validating the design procedure. Compared to a Gaussian profile of identical $w_0$, also shown, the super-Gaussian mode remains constant for most of the resonator volume but then decreases much faster towards the mirror edges, minimizing diffractive losses. 
The resulting $\mathrm{FoM}=0.57$, which is more than double the maximum FoM of the corresponding spherical mirror design.
This is largely due to an increased form factor, $C=0.57$, while the $Q$-factor remains similar, $Q_0=57500$. 
We denote this configuration as the \emph{cryo, graded} configuration in Table~\ref{tab:configs}.
Following Eq.~\eqref{eq:scan_rate},
this would lead to a factor $\sim 4.4$ speedup in scan rate.
The theoretically maximum form factor for a truly uniform mode in transverse direction and a $\cos(2\pi z/\lambda)$ dependence in axial direction is $8/\pi^2 \approx 0.8$. Consequently, using even higher-order super-Gaussians with $N>3$ does not appear worthwhile as this would require an increasingly complex mirror profile for relatively small further improvements in form factor.
We show the degradation of the signal power when scanning over frequency for a fixed set of graded-phase mirrors in Fig.~\ref{fig:scan_range}.

\section{Discussion and Conclusions}
\label{sec:conclusions}

In this work, we have assessed the sensitivity of a Fabry-P\'erot resonator, coined \faxe, to the detection of dark matter axions at frequencies above 30\,GHz. 
Even with comparably simple and cost-effective configurations, 
the sensitivity of \faxe surpasses current constraints on the photon-ALP coupling at masses $m_a \gtrsim \SI{0.1}{\milli\eV}$, cf. Fig.~\ref{fig:m-vs-g}. 
For our baseline cryogenic benchmark configuration (see Table~\ref{tab:configs}) employing spherical mirrors, one should be able to probe values of $g_{a\gamma}$ predicted by  specific axion dark matter scenarios \cite{2021JHEP...01..172C,2021arXiv210201082D}.
If graded-phase mirrors are used instead, the mode profile becomes super-Gaussian, which  leads to an improvement of the form factor by roughly a factor of 2.

Concerning the required alignment precision, we expect relative tilts to be of secondary concern owing to the fact that the separation between mirrors is much smaller than their diameter. Consequently, rather large tilts of order $\mathcal{O}(\si{\milli\radian})$ are required to move the resonating eigenmode considerably off-center resulting in increased clipping losses. Deviations from an ideal mirror surface lead to scattering losses which can be estimated by Ruze's equation~\cite{ruze}. To not be limited by scattering losses, we require the RMS deviations to be smaller than $\SI{5}{\micro \meter}$ for the cryogenic benchmark.

To reach a sensitivity to probe QCD axion benchmark models requires an upscaled version of \faxe, which we refer to as \faxe \emph{strong}. 
In particular, we can improve the sensitivity with (i) a higher magnetic field, (ii) larger mirrors, and (iii) a higher quality factor of the resonator. 
Points (i) and (ii) could be addressed with a dedicated experimental platform, such as the  planned Dark Wave lab at Fermi Lab~\cite{ADMX-EFR}. 
It features a repurposed MRI magnet with a field strength of $\SI{9.4}{\tesla}$ with a $\SI{800}{\milli\meter}$ bore and a cryostat to reach temperatures of $\SI{4}{\kelvin}$ and potentially $\SI{100}{\milli\kelvin}$.
Regarding (iii), the graded-phase mirrors could be coated with superconducting (SC) material. 
With such mirrors, an FPR with a finesse of $\mathcal{F} = 4.6\times10^9$ and quality factor of $Q_0 = 4.2\times10^{10}$ at a frequency of 51\,GHz has been achieved~\cite{Kuhr2007}.
The mirrors with a radius of $r=\SI{25}{\milli\meter}$ and separation of $\ell \approx \SI{28}{\milli\meter}$ were coated with a $\SI{12}{\micro\meter}$ layer of niobium. 
For an application in \faxe, a type~II superconductor would be necessary that features a high value of the critical magnetic field, $B_c \gtrsim \SI{10}{\tesla}$ for a $B$ field parallel to the mirror surface and a critical temperature above $\sim\SI{4}{\kelvin}$. 
Potential options are Nb$_3$Sn~\cite{2017SuScT..30i3001X} or the high temperature superconductor REBCO as used in the RADES haloscope~\cite{2022ITAS...3247741G}.

Clearly, manufacturing such large mirrors to high precision and coating them uniformly with a superconductor is a technical challenge that will require further investigation.
It should also be noted that for such large mirrors with $r=\SI{40}{\centi\meter}$, the critical coupling via an aperture leads to distortions of the fundamental mode. 
A detailed investigation is left for future studies but first numerical results indicate only minor deterioration.
Additionally, mechanical alignment will need to be precisely controlled to avoid mode localization. 

Motivated by the results obtained in Ref.~\cite{Kuhr2007,2022ITAS...3247741G}, we use a quality factor of $Q_L = 10^6$, so that $Q_L = Q_a$.
Lastly, we assume a system noise temperature equal to twice the standard quantum limit~\cite[e.g.,][]{2018arXiv180100835B}, $T_\mathrm{sys} = 2hf/k_B\sim\SI{6.7}{\kelvin}$ at $f=\SI{70}{\giga\hertz}$ (the highest scan frequency we consider here), which would require near quantum-noise-limited amplifiers.
We summarize all the above parameters for \faxe strong in Table~\ref{tab:configs} and show the final sensitivity in Fig.~\ref{fig:m-vs-g}.
With a world-record volume of $2500\lambda^3$ to $14000\lambda^3$ (depending on frequency), \faxe \emph{strong} could probe the KVSZ and DFSZ QCD axion 
benchmarks with a 
total observations time of $\sim 9.6$~years corresponding to a scan rate of $\sim4.2\unit{GHz.yr^{-1}}$. 
For comparison, the 1B~run of ADMX had a scan rate of $157\,\unit{MHz.yr^{-1}}$ to reach the sensitivity required to probe the DSFZ axion~\cite{2020arXiv201006183A}. 
The CAPP haloscope experiment at CAPP reached $\sim500\unit{MHz.yr^{-1}}$ in their first DFSZ run~\cite{2023PhRvL.130g1002Y} and even faster rates up to $\sim0.8\unit{GHz.yr^{-1}}$ for subsequent data taking campaigns~\cite{2024PhRvX..14c1023A}.  
This highlights once more that \faxe could scan the frequency range relatively fast due to its unique combination of large effective volume and high quality resonance.

We compare the sensitivity of \faxe strong with future sensitivities of other haloscope experiments in Fig.~\ref{fig:m-vs-g}.
\faxe will overlap with the high-frequency regions of ALPHA, MADMAX and ORGAN. 
As mentioned in Sec.~\ref{section:Intro}, ORGAN will operate a number of different cavities to scan various axion masses. 
The planned phases D-G from ORGAN will overlap with the \faxe frequency range. For the sensitivity projections, it is assumed that the goal of quantum limited amplification is met over the course of the experiment~\cite{2017PDU....18...67M}. 
MADMAX, on the other hand, is a dielectric haloscope, in which the form factor is effectively improved with eventually 80 dielectric disks with a diameter of $\SI{1.25}{\meter}$, whose positions need to be carefully tuned to the axion mass. 
The experiment will also require a dedicated magnet with a planned field strength of $\SI{9}{\tesla}$ and a large bore to fit the disks~\cite{2020arXiv200310894B}.
Instead of matching resonance frequency of cavity to axion mass by changing cavity dimensions, ALPHA aims to match the effective plasma frequency with the mass of the axion thereby achieving resonance~\cite{2019PhRvL.123n1802L}. 
Wire metamaterials provide the effective plasma frequency and tuning is achieved through changing the distance between (pairs of) wires~\cite{2023PhRvD.107e5013M}.
Special attention will be required to avoid mode hoping between TE and TM modes.
Projections assume a field strength of $\SI{13}{\tesla}$ and a bore radius of $\SI{35}{\centi\meter}$.
The indicated axion parameter space could be scanned within 10 years of observation time with different inserts.

It is clear that the different planned experiments will face different technological challenges (also noted in, e.g., Ref.~\cite{2023PhRvD.107e5013M}). 
\faxe will be a complementary probe for axion dark matter that will help to probe this challenging frequency region.

\begin{acknowledgments}
The authors would like to thank the anonymous referees for helpful comments and suggestions.
They also thank Elmeri Rivasto for discussions on high temperature superconductors. M.M. acknowledges support from the European Research Council (ERC) under the European Union’s Horizon2020 research and innovation program Grant Agreement No. 948689 (AxionDM). J.E. and M.M. also acknowledge support from the Deutsche Forschungsgemeinschaft (DFG, German Research Foundation) under Germany’s Excellence Strategy – EXC 2121 “Quantum Universe”– 390833306. This article is based upon work from COST Action COSMIC WISPers CA21106, supported by the European Cooperation in Science and Technology.
\end{acknowledgments}

\section*{Data Availability}
The data that support the findings of this article are openly available \cite{dataset_zenodo}.

\bibliography{mainbib} 


\end{document}